\documentclass[sigconf]{acmart}

\usepackage{amsmath,amssymb,amsfonts}
\usepackage{adjustbox}
\usepackage{listings}
\usepackage{color,soul}
\usepackage{pifont}
\usepackage{tcolorbox}
\usepackage[htt]{hyphenat}

\usepackage{blindtext}

\usepackage{wrapfig}
\usepackage{tikz}
\usetikzlibrary{decorations}
\usetikzlibrary{backgrounds}
\usetikzlibrary{arrows,shapes}
\usetikzlibrary{tikzmark}
\usetikzlibrary{calc}

\definecolor{dkgreen}{rgb}{0,0.6,0}
\definecolor{gray}{rgb}{0.5,0.5,0.5}
\definecolor{mauve}{rgb}{0.58,0,0.80}

\renewcommand\footnotetextcopyrightpermission[1]{} 
\usepackage{booktabs} 
\usepackage[shortlabels]{enumitem}

\setcopyright{none}
\usepackage{titlesec}
\begin{document}

\titlespacing*{\section}{0pt} {0.5\baselineskip}{0.5\baselineskip}
\titlespacing*{\subsection} {0pt} {0.5\baselineskip}{0.5\baselineskip}

\captionsetup{skip=0pt}

\title{A Machine Learning-Based Approach For Detecting Malicious PyPI Packages  }


\author{Haya Samaana$^1$, Diego Elias Costa$^2$, Emad Shihab$^2$, Ahmad Abdellatif$^3$}
\affiliation{%
  \institution{$^1$An Najah National University, Nablus, Palestine; hayasam@najah.edu \\
               $^2$Concordia University, Montreal, Quebec, Canada; diego.costa@concordia.ca, emad.shihab@concordia.ca \\
               $^3$University of Calgary, Calgary, Alberta, Canada; ahmad.abdellatif@ucalgary.ca}
}

\begin{abstract} 


\textbf{Background}. In modern software development, the use of external libraries and packages is increasingly prevalent, streamlining the software development process and enabling developers to deploy feature-rich systems with little coding. 
While this reliance on reusing code offers substantial benefits, it also introduces serious risks for deployed software in the form of malicious packages - harmful and vulnerable code disguised as useful libraries.
\textbf{Aims}. Popular ecosystems, such PyPI, receive thousands of new package contributions every week, and distinguishing safe contributions from harmful ones presents a significant challenge. 
There is a dire need for reliable methods to detect and address the presence of malicious packages in these environments. \textbf{Method}. To address these challenges, we propose a data-driven approach that uses machine learning and static analysis to examine the package's metadata, code, files, and textual characteristics to identify malicious packages. \textbf{Results}. In evaluations conducted within the PyPI ecosystem, we achieved an F1-measure of 0.94 for identifying malicious packages using a stacking ensemble classifier. \textbf{Conclusions}. This tool can be seamlessly integrated into package vetting pipelines and has the capability to flag entire packages, not just malicious function calls. This enhancement strengthens security measures and reduces the manual workload for developers and registry maintainers, thereby contributing to the overall integrity of the ecosystem.

\end{abstract}






\maketitle

\section{Introduction} \label{intro}



Code reuse drives technological innovation by enhancing developer productivity and enabling the creation of feature-rich, maintainable systems \cite{mohagheghi2004empirical,basili1996reuse,grinter1996understanding}. Package managers like npm and PyPI facilitate this innovation, with PyPI allowing developers to contribute freely to a vast repository of Python packages \cite{kaplan2021survey,vu2020towards}. PyPI meets diverse needs, including those in artificial intelligence, with millions of packages available \cite{liang2021malicious}. Its popularity has made Python the most favored programming language as of April 2024, according to the TIOBE index.
\footnote{https://www.tiobe.com/tiobe-index/}


Developers are increasingly reusing code, which inadvertently heightens the risk of integrating malicious code into applications \cite{ladisa2023sok}. Malicious actions can include inserting backdoors and stealing sensitive data. The PyPI registry faces various attacks, including compromised accounts, where attackers gain full control of a maintainer's account to publish malware \cite{vu2020typosquatting, kaplan2021survey}. Typosquatting \cite{bertusk, dateutil_typo} exploits typographical errors in package names, while combosquatting \cite{vu2020typosquatting,tschacher2016typosquatting} leverages the order of nouns in names. A notable example is the malicious package "jeIlyfsh," which replaced a character in the benign "jellyfsh" package name to steal SSH and GPG keys, remaining undetected for a year \cite{vu2020towards}.    

Several code-scanning methods have been proposed to identify malicious packages in popular ecosystems like NPM, PyPI, and Maven. These approaches encompass both traditional methods (e.g., anomaly detection \cite{liang2021malicious}, dynamic and static analysis \cite {duan2020measuring, duan2020towards}, and machine learning-based methods (e.g., unsupervised k-means clustering \cite{garrett2019detecting}, supervised learning \cite{sejfia2022practical}). 
However, managers of large package registries, such as PyPI, struggle to handle the daily influx of packages, making it challenging for manual oversight \cite {zhang2023malicious, gu2023investigating}. According to the libraries.io database \cite{Libraries}, which is an open source repository and dependency database that catalogues libraries of the most popular ecosystems, and it has been used by previous work as a source of library metadata \cite{decan2018impact,alfadel2021empirical}, analysis reveals that  over the course of one week, PyPI developers publish around 1,800 public package versions, including both new packages and updated versions of existing packages. 
To this end, prior works show that many existing scanning tools face limitations such as scalability issues \cite{sejfia2022practical, maloss}, a narrow focus on specific ecosystem aspects like package updates \cite{sejfia2022practical, garrett2019detecting}, high resource costs \cite{maloss}, and high false alerts \cite {bandit, maloss}.   
In this context, we have detailed code scanners that
work well if we analyse a few packages, however, there is a necessity for an approach that can extend its scalability to encompass the entire ecosystem. Our approach is designed to fill this gap, addressing issues of false positives and negatives. Its strength lies in the integration of newly crafted features operating at the package level, distinguishing it from most existing tools that operate at the function call level.
Our approach considers the interactions of multiple factors across code, function, file levels, and package metadata, to provide a scalable classification of the likelihood of a malicious package. 
To evaluate the effectiveness of our approach, we conduct a study to answer the following three research questions.\\
\noindent
\textbf{ RQ1: How accurate is our approach in classifying malicious packages?}
We developed six machine learning classifiers—Random Forest, Decision Tree, Support Vector Machine, Multilayer Perceptron, Naive Bayes (Bernoulli version), and stacking—using eight features: 2 metadata-related, 2 file-related, 3 code-related, and 1 text feature. The classifiers were evaluated on a dataset of 5,193 benign and 138 malicious packages. Additionally, we tested the method's generalizability on an unseen dataset of 397 typical and 143 new malicious packages. The stacking ensemble classifier achieved an F1-score of 94.2\% in predicting malicious packages, particularly when incorporating the text-related feature. Our findings indicate strong performance and generalizability of the approach.
 
\noindent\textbf{ RQ2: What features are the best indicators of malicious
packages?}
The study investigates features that differentiate malicious packages from benign ones, identifying metadata-related features as the best indicators. Key tokens like \textsc{getattr, connect, read, open} are crucial for malicious package identification. The research also examines the effects of keyword removal, stop word removal, and stemming on classifier performance. It concludes that stemming does not influence performance, while removing keywords and stop words has minimal impact.
   
\noindent
\textbf{RQ3: Is our approach useful?}
We assess the viability and efficiency of our approach by comparing it with two state of the art tools, namely bandit and packj. This evaluation is conducted on a random subset comprising 50 benign packages and an additional 50 samples of malicious packages obtained from an external dataset and is performed on two scenarios: the entire package and the setup.py file. 
The results suggest that our approach effectively detects a considerable number of malicious packages in real-world scenarios, demonstrating a low rate of false alerts when compared to the tested tools. As an illustration, our tool successfully detected 4 out of 5 recently released malicious packages, outperforming the tested tools that proved unsuccessful in this regard. Furthermore, by selecting features related to the Python ecosystem, our method can detect a broader spectrum of malicious Python packages, in contrast to the state-of-the-art approach~\cite{ohm2022feasibility}.

Our study contributes to the research and practice on four fronts.
\vspace{-\topsep}
\begin{itemize}
\item Unlike existing strategies that focus on detecting suspicious function calls, our approach operates at the package level, applicable to the entire ecosystem.
\item This is the first study to employ a vocabulary-based method to automatically detect malicious packages within the PyPI ecosystem.
\item We introduce significant new features previously unexplored in predicting malicious functions and packages.
\item We publicly share our dataset \cite {dataset_link} for further research on enhancing the security of package managers against supply chain attacks, comprising 5,331 packages (138 malicious and 5,193 popular) with various features.
\end{itemize}

\section{Background}\label{malicious}

Malicious packages are software components intentionally designed to harm systems, often containing malware or code for unauthorized activities like stealing sensitive information or installing backdoors \cite{stealing, cryptocurrency}. In contrast, benign packages are safe and intended to perform their designated functions without compromising security.

Numerous studies indicate that malware is continuously evolving, employing diverse techniques to evade detection tools \cite{duan2020towards,xu2013jstill,xu2012power}. Attackers target all stakeholders in the software supply chain, including end-users, developers, package maintainers, and registry maintainers. One prevalent tactic is typosquatting and combosquatting \cite{tschacher2016typosquatting, vu2020typosquatting}, as many registries lack security policies \cite{duan2020towards}. In typosquatting, malicious packages are published with names similar to popular packages to deceive developers into downloading them. Combosquatting manipulates the order of words in package names, such as changing 'python-nmap' to 'nmap-python.' An illustrative incident occurred in May 2022 when attackers published a malicious package named pymafka, which mimicked the legitimate PyKafka package, leading to 325 downloads before its removal \cite{pymafka}. Additionally, attackers publish new malicious packages directly, often employing code obfuscation methods to conceal harmful code from analysis. Techniques like encoding and encryption are commonly used, as seen in the colourama@0.1.6 typosquatting variant of colorama, which utilized base64 encoding to evade detection as shown in listing \ref{colourama} (Line 1). Malicious packages like hipid and hpid \cite{encode32}  have been reported using uncommon base32 encoding, while the botaa3 typosquatting package employs bitwise XOR encryption and base64 encoding to obscure its malicious payloads \cite{botaa3}.

Many tools are developed to detect malicious attacks against popular ecosystems (e.g., NPM and PyPI) such as Malware-check  \cite{malware-check-tool}, OSSGadget \cite{OSSGadget}, Maloss \cite{maloss}, Packj \cite{packj-vetting-tool}, Bandit4mal \cite{Snakesndit4malPyPI}, and Bandit \cite{bandit}.
To the best of our knowledge, the proposed tools work in detecting risky factors in Python packages, always reporting alerts at the level of Python functions. To put our results into perspective, we opt to select both Packj and Bandit as baselines in our evaluation because 1) they have been selected as benchmarks in previous research \cite{vu2020towards, vu2022benchmark, sejfia2022practical} and 2) they are mature tools that provide detailed reports that facilitate our manual analysis.
The \textbf{Bandit} tool \cite{bandit} is a widely-used static analysis tool designed for identifying security vulnerabilities in Python files. It employs predefined rules and the Abstract Syntax Tree (AST) representation of source code to enhance its analysis. Bandit rates issues based on severity and confidence levels (low, medium, high), providing insights into potential impact and reliability  \cite{ruohonen2021large}. It has also been used as a benchmark in previous research \cite {vu2022benchmark,vu2021lastpymile}.
On the other hand, the \textbf{Packj} tool \cite{packj-vetting-tool}  employs a comprehensive analysis approach with three steps: static code analysis, metadata analysis, and optional dynamic analysis. It examines package code for filesystem, network, and process API usage, validates metadata attributes, and can perform dynamic analysis. Packj is built upon the MalOSS project \cite{maloss} and is recognized in research as a benchmarking framework \cite{vu2020towards, vu2022benchmark, sejfia2022practical}.
Given that attackers employ diverse tactics to obfuscate the detection of malicious packages, and existing registries lack a robust review process for package publication \cite{duan2020towards}, and most existing tools suffer from considerable limitations, our emphasis is on integrating various features. These include meta-related, code-related, and text-related features. These features aim to effectively capture the range of tactics employed by attackers, enhancing the ability to identify malicious packages.

\begin{lstlisting}[caption={colourama package \cite{colourama} uses code obfuscation in setup.py file to defeat analysis (snippet code).},language=Python,label={colourama}]
def run(self):
    exec("b3MxID0WNv...cmludA==".decode('base64')) \\Line 1
    os = platform.system()
    req = urllib2.Request('https://grabify.link/...', 
    texto = urllib2.urlopen( req ).read() 
\end{lstlisting}

\section{Related work} \label{related_work}


\par
\noindent
\textbf{Traditional approaches.}
Several methods have been proposed for identifying malicious packages. \citet{liang2021malicious} used anomaly detection, combining abstract syntax tree (AST) and regular expression techniques, achieving a 97.51\% reduction in review workload, though it struggles with small malicious artifacts. \citet{duan2020measuring} reported 339 malicious packages through dynamic and static analysis but lacked false positive analysis. Their MALOSS framework, while comprehensive, requires significant resources. \citet{vu2021lastpymile} identified differences between published packages and source repositories but missed packages without repositories. \citet{ohm2020towards} relied on dynamic analysis with heavy manual intervention, while \citet{rieck2011automatic} monitored behavior in a sandbox but lacked detailed analysis. Various methods targeting typosquatting often suffer from false positives and negatives due to their focus on Levenshtein distance alone \cite{taylor2020spellbound, vu2020typosquatting, tschacher2016typosquatting}.
\begin{table}
\vspace{-4mm}
\caption{Existing techniques for analyzing PyPI suspicious/malicious packages.}
    \centering
    \begin{adjustbox}{width=\columnwidth}
    \small
    \huge
\begin{tabular}{lll}
     \toprule
     \textbf{Tool}&\textbf{Input}&\textbf{Technique} \\
     \midrule
     
    Malware Checks \cite{malware-check-tool} & setup.py file & static (Regular Expression)\\
    OSSGadget  \cite{OSSGadget} & package+artifacts &  static (Regular Expression)  \\
    
     MalOSS \cite{duan2020towards} & package & hybrid (metadata, static, dynamic)  \\
    Packj \cite{packj-vetting-tool} & package & hybrid (metadata, static, dynamic)  \\

    bandit4mal \cite{Snakesndit4malPyPI} & package & static (Abstract Syntax Tree)  \\

     bandit\cite{bandit} & package & static (Abstract Syntax Tree)  \\
    \hline

\end{tabular}
\small


    \end{adjustbox}

    \label{tab:scanning techniques}
    \vspace{-4mm}
\end{table}

\noindent
\textbf{Machine learning approaches.}
Related work includes \citet{garrett2019detecting}, who used unsupervised learning to identify suspicious NPM package updates based on features like resource access and API usage, flagging 539 updates but not investigating malicious packages in depth. \citet{sejfia2022practical} proposed an automated method for detecting malicious NPM packages that examined version changes and employed classifiers, successfully identifying 95 out of 96,287 unknown malicious packages with acceptable false positives. However, their approach lacks scalability and is limited to package updates, unable to address single-release packages.
\citet{halder2024malicious} developed MeMPtec, which utilizes features from package metadata. However, our approach goes deeper by analyzing metadata and file-related features. We parse the contents of license and configuration files, rather than simply reporting their existence.
Recent research by \citet{ohm2022feasibility} focuses on detecting malicious NPM packages using a combination of classifiers, achieving true positive rates over 70\% by evaluating 25,210 models. Their optimal combination involved Support Vector Machine, Multi Layer Perceptron, and Random Forest, successfully identifying 13 previously unknown malicious packages. This work is the most closely related to ours, as it also seeks to classify malicious packages within the NPM ecosystem. However, our methodology differs significantly. We adopt a more holistic approach by incorporating features related to code, licenses, file configurations, and author information. Unlike \citet{ohm2022feasibility}, which treats suspicious APIs as boolean features, we analyze entire lines of code as text features. To our knowledge, no previous research has integrated linter outputs with metadata to classify software packages as malicious.
\citet{ohm2022feasibility} developed a methodology tailored for the NPM ecosystem, specifically focusing on identifying malicious JavaScript packages with features relevant to that environment. In contrast, our work is explicitly designed for the Python ecosystem. We have crafted features that incorporate python-specific setup configuration files, allowing for a more accurate analysis of potential threats within Python packages. This distinction highlights the adaptability of security measures to different programming ecosystems.

\noindent
\textbf{Python malware detection tools.}
\label {tools} In the literature,  a multitude of methods such as malware-check  \cite{malware-check-tool}, OSSGadget \cite{OSSGadget}, Maloss \cite{maloss}, Packj \cite{packj-vetting-tool}, Bandit4mal \cite{Snakesndit4malPyPI}, and bandit \cite{bandit} have been suggested for detecting suspicious package. The majority of these methods examine various facets of a package through the utilization of metadata, static analysis, or dynamic analysis as shown in Table \ref{tab:scanning techniques}. Analysis methods for metadata (such as package name and author information) involve scrutinizing these metadata elements to detect potentially problematic packages. One instance of this is the utilization of package names and their popularity to identify suspicious packages that might involve tactics like typosquatting or combosquatting.

\begin{figure}
\includegraphics[width=\columnwidth]{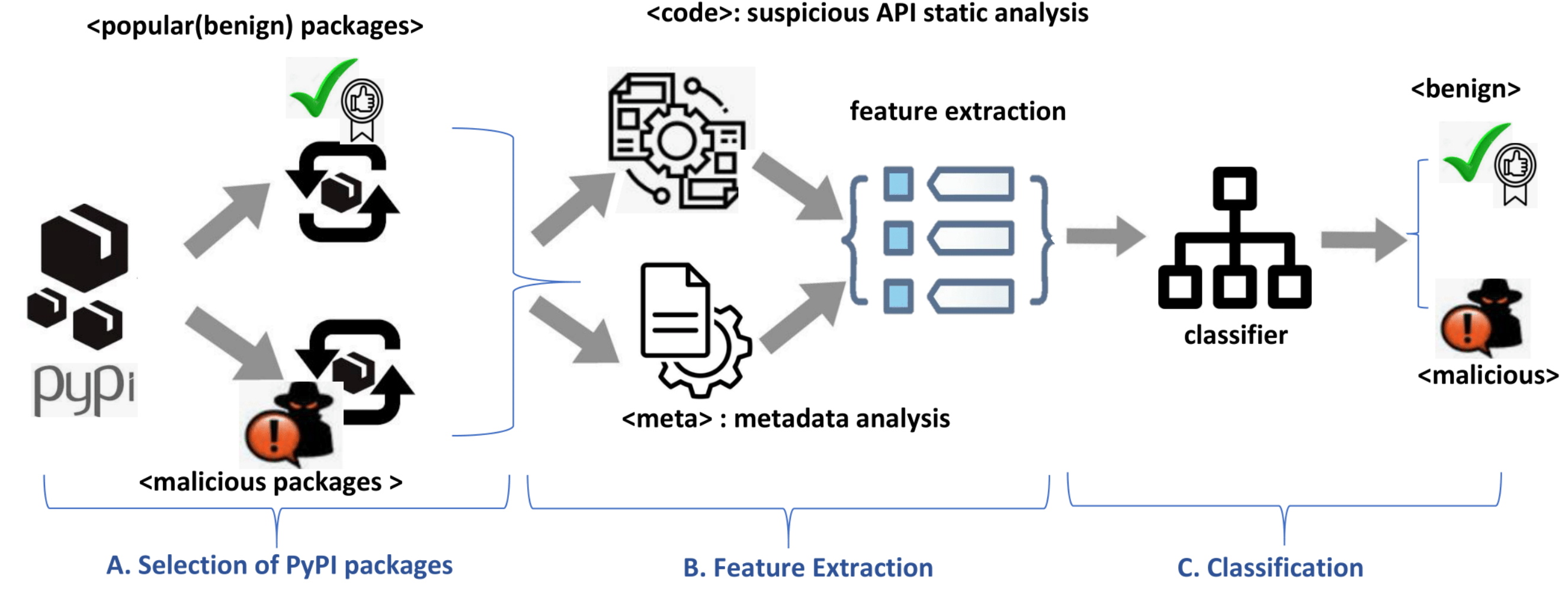}

\caption{The workflow for identifying malicious packages.}
\vspace{-6mm}
\label{fig:workflow}
\end{figure}

However, existing tools struggle with unproductive time use, overlooked security issues, and high resource demands for dynamic analysis. Many, like OSSGadget and malware-check, rely on limited rule-based detection, potentially missing harmful code and hindering efficient detection of malicious actions.

\noindent
Unlike previous research efforts that created comparative frameworks for different ecosystems or aimed at identifying suspicious packages at the function calls level, our research stands out by focusing on identifying malicious packages at the package level. Notably, it surpasses the scope of proposed works by evaluating six classifiers on a comprehensive set of newly crafted features and features inherited from existing literature, ensuring the successful detection of malicious packages throughout the entire ecosystem.

\section{Study Setup}\label{study_setup}
The main goal of our study is to identify malicious packages from a set of suspicious packages. 
Many techniques are proposed for this purpose as shown in Table \ref{tab:scanning techniques}, however, they do not work at the package level, and requiring maintainers to inspect multiple alerts before identifying malicious packages. 
Therefore, in order to attain our objective and address the limitations of current methods, we develop a machine learning (ML) approach. This approach is centered around combining specific static features, as demonstrated in the upcoming Section \ref{sub:feature_selection}. We resort to use ML technique as it is a popular technique in the area of information security \cite{hou2010malicious,duan2020towards,ohm2022feasibility,sejfia2022practical}.

For a comprehensive view, Figure \ref{fig:workflow}
shows the full workflow of our approach, which consists of three
phases: collecting benign and malicious PyPI packagess (Section \ref{sub:collecting-PyPI-libraries}), feature extraction including metadata, file, code, and text related features (Section \ref{sub:feature_selection}), and classification phase which details the machine learning classifiers and presents the evaluation process (Section \ref{sub:evaluation}).


\begin{table*}

    \caption{ An overview of feature set used to identify malicious packages in PyPI, including reused features from \citet{ohm2022feasibility}.}
      
    \small \begin{adjustbox}{width=\textwidth}
\begin{tabular}{lll|c}
      
     \toprule
     \textbf{Dimensions}&\textbf{Features}&\textbf{Definition}&\textbf{Reused Features From \citet{ohm2022feasibility}}   \\
     \midrule

    \textbf{Metadata-related} & Has invalid or no homepage & The package includes invalid or no homepage. &\ding{55}\\
                              &  Has invalid or no author email & The package includes invalid or no author email.&\ding{55}\\
             \hline
     \textbf{Text-related}              & Suspicious LOC & Lines of code including suspicious APIs.& \checkmark \\

    & & & (Consider each API as a boolean feature) \\

     \hline
       
     \textbf{Code-related}  & Has (post) install command & The package includes install script.&\checkmark\\
     
                      & Has suspicious URL & The package includes suspicious URL or IP address.&\checkmark \\
                      & Has long string & The package includes very long string (obfuscation).&\checkmark\\
                     
                      \hline
     \textbf{File-related} 
                      & Has minimum setup configuration   & The  developer does not  specify the package details in the setup.cfg file.&\ding{55} \\
                      & Has mismatch license   & The package is missing license type uniformity in the three core positions. &\ding{55}\\
     \hline
 
     \hline
     
      \end{tabular}
 \end{adjustbox}

     \vspace{-4mm}
    \label{tab:features}
\end{table*}

\subsection{\textbf{Collecting Training and Testing Datasets}}
\label{sub:collecting-PyPI-libraries}


\noindent
\textbf{Training dataset.}
One of the main challenges in building a classifier model to identify malicious packages is the quality and abundance of data.
As thousands of packages are published weekly, it is well known that only a small fraction of malicious packages are immediately flagged as security threats; hence, we cannot assume that other recently published packages are benign (non-malicious).  
To solve this problem, we use popular packages that have been used by projects for years to compose the training/validation set of non-malicious packages, similarly as done in prior work \cite{ohm2022feasibility, zahan2022weak,taylor2020spellbound}. 
Also, widely used and trusted popular packages are often copied by malicious packages (e.g., typosquatting), aiming to camouflage their activities and reach a larger number of users and developers \cite{gu2023investigating, ohm2020backstabber}.
As such, our approach must be able to distinguish between popular packages and their malicious copies. 
It is important, however, to note that 1) we refrain from using any popularity metric as a feature of our classifier, our classifier should use only code and metadata-related features, and 2) we also test our model's performance on a set of benign packages that are also not popular, to simulate a real-case scenario of using our approach.  

To build our training and validation dataset, we collect \textbf{5,193 benign packages}.
We first collect the top 5,000 most downloaded PyPI packages, from the PyPI registry~\cite{hugo_van_kemenade_2022_popular} and select the top 5,000 most dependent upon packages from PyPI, as recorded in the libraries.io database~\cite{Libraries}.
We then merge both datasets, removing all duplicates (the vast majority), leading to a total of 5,193 popular Python packages in the PyPI ecosystem.





To build our set of malicious packages for model training, we use the dataset of 252 malicious packages collected by \citet{ohm2020backstabber}. 
As this dataset contains multiple versions of the same malicious packages, we only kept the latest version of malicious packages, similarly as done in previous work \cite{liang2021malicious}, to avoid overrepresenting a single package in out training set. We also remove (18) packages that were deemed not complete (e.g., package that included only the setup.py file or just the malicious payload). Finally, the malicious dataset contains (138) packages.
This dataset of malicious packages spans 2015 to 2023, and the majority of attack vectors target install time rather than runtime.
The dataset includes a variety of different injection techniques, such as TypoSquatting (52\%) and Trojan Horse (27\%), with varying infiltration objectives, e.g., data exfiltration (44\%), droppers (15\%), backdoors (8\%). 
More information can be found in the study of ~\citet{ohm2020backstabber}.
\\

\noindent
\textbf{Test dataset.}
While using popular packages as a proxy for benign packages is a sensible choice, it is expected that distinguishing between popular packages and malicious packages is easier than finding malicious packages in a batch of newly published Python packages~\cite{vu2023bad}.
Thus, we craft a test dataset that, by definition, does not contain any package (malicious or not) seen by the model during training but also better represents the average packages in the PyPI ecosystem.
To achieve this, we 1) randomly select 397 packages from the PyPI registry, hereby named as \textbf{typical packages}. 
For the malicious packages, we incorporated a new collection of \textbf{143 malicious packages}, distinct from our training dataset selected from the dataset introduced by \citet{ohm2020backstabber}. During the implementation of the study, the dataset maintainer supplemented the collected training dataset with these 143 malicious packages, which we subsequently designated as the test dataset.


   

\subsection{Feature Extraction}
\label{sub:feature_selection}
We need to collect features that capture different characteristics of malicious packages. We observed that majority of related work targeted NPM ecosystem, thus a set of features are adapted from NPM ecosystem \cite{sejfia2022practical,garrett2019detecting}, called Code-related features in our case. Other features were chosen
based on the grey literature \cite{licenses} and expert knowledge of the differences between benign and malicious packages (in our case called File-related features). Moreover, a few of
metadata-related features ( e.g., README length, dependency analysis, author information, homepage, number of versions) are explored in the context of machine learning based solution \cite{liang2021malicious, vu2021lastpymile, sejfia2022practical}. In general, it's important to note that not all metadata features hold significance; certain attributes might introduce noise and subsequently impact the model's performance. Hence, we expand the feature set to include metadata-related features. Finally, a text-related feature was derived from various sensitive APIs and permissions included in the source code. These APIs have the following behaviours: produce new code during runtime ("getattr"), create forks or terminate operating system processes ("exit", "Thread", "system"), access obfuscated (hidden) code, modify system or environment variables ("clear"), access files and directories ("open"), establish connections with external networks ("open\textunderscore connection"), or read\/write user input ("input", "mkdtemp").

Our approach involves integrating eight different sets of features summarized in Table \ref {tab:features}, to capitalize on the synergies that can arise
from the interactions among these distinct features.
\\
\textbf{Metadata-related features.} Every PyPI package has a set of metadata that contains different information such as homepage, project-url, authors and maintainer-email. 
For the computation of these features, we analyze the PKG-INFO file associated with each package. The validity of the email address is assessed by confirming its domain, while the homepage URL's validity is verified by ensuring its security and association with a well-known domain. The compilation of popular domains is derived from the list provided in \cite{domains}.
We consider two features of this dimension in our approach. \\
\noindent
\underline{\textit{Has invalid or no homepage (repository):}} Previous work showed that the differences in source code between build artifacts of a package and the respective source code repository are a strong indicator for its maliciousness~\cite {vu2021lastpymile, zahan2022weak}.
Our hypothesis suggests that this feature could aid in differentiating between malicious and benign packages, as supported by previous studies that emphasize the importance of having a homepage or/and source code repository for a given package. Moreover, based on a quantitative analysis of our dataset, we find that only 20\% of malicious packages have a valid homepage or repository, while 73\% of benign packages own a valid homepage. \\
\underline{\textit{Has invalid or no author email:}}
 A recent investigations highlight that packages lacking a valid author email address, often a result of improper packaging guidelines, may raise suspicion and serve as an indicator of potential malicious intent \cite{vu2023bad, zimmermann2019small}. Among the inspected packages of the before mentioned dataset, we find about  87\% of benign packages have a valid author's email address in comparison to only 46\% valid email address in malicious packages. 
 
\begin{lstlisting}[caption={Snapshot of a generated report from packj tool},language=Python,label={packj-report}]
"performs a process operation": [{
    "filepath": "10Cent10-999.0.4\\setup.py",
    "api_name": "spawn",
    "lineno": "17"
}]
\end{lstlisting}

\textbf{Text-related feature:}
Most known attacks had malicious code injected into setup.py file \cite{ohm2020backstabber, vu2020towards}. 
In this particular situation, our hypothesis revolves around the possibility that the malicious attacker might disperse the harmful payload among various files. To better understand and address this scenario, we extract lines of code corresponding to the suspicious APIs. To avoid reinventing the wheel, we rely on the generated static analysis report of the packj tool \cite{packj-vetting-tool} to construct a portion of text feature. These APIs are linked to file paths and line numbers, as demonstrated in listing \ref{packj-report}. 
 To formulate a complete text feature, we enrich these suspicious lines of code by including the source code that originates from the setup.py file, in addition to any other .py source code that involve suspicious URLs.
Moreover, based on a quantitative analysis of our dataset, we observe that setup.py file was the most targeted file by attackers (75\%). Our observation is inline with prior works \cite{ohm2020backstabber, vu2021lastpymile, Snakesndit4malPyPI} showed that setup.py file is the most likely file to be manipulated by attackers. 

\textbf{Code-related features.} Many characteristics have been extensively discussed in prior research studies like \cite{sejfia2022practical, ohm2022feasibility, liang2021malicious, garrett2019detecting}, highlighting their significance in proficiently recognizing malicious software packages. These characteristics include installation command, suspicious URL, and long string. In this context, we leverage the previously discussed generated text feature to extract the following features.\\
\underline{\textit{Has (post) install command:}} Prior works \cite{ohm2022feasibility,sejfia2022practical} show that the install command initiates an external operation, which is a common characteristic observed in malicious packages. Thus, we use regular expressions to search for the install command in the generated text feature.\\
\underline{\textit{Has suspicious URL:}}
Different studies showed that attackers often inject their IP address or URL address in malicious code \cite{liang2021malicious,ohm2022feasibility}. We use regular expression to extract URLs and IP addresses for a particular package, then we record the suspicion of each URL/IP based on different criteria such if the URL is insecure, and if it belongs to unpopular domain. We rely on a list of approved URL domains provided by Amazon to verify the domain of a URL. This list contains the top 1 million most popular domains from Alexa, and if the domain of the URL is not present in the list we mark the URL as suspicious \cite{domains}.\\
\underline{\textit{Has long string:}} The obfuscated code is often very long \cite{kim2011suspicious, liang2021malicious}. Attackers commonly apply specialized encoding methods such as base64 to obscure harmful payloads. A string is categorized as lengthy when its length surpasses a specific threshold (40 characters) \cite{canali2011prophiler}.
Consequently, we adopt this characteristic with string length larger than 40, following prior works \cite{liang2021malicious,ohm2022feasibility, canali2011prophiler}. Our dataset indicates that this attribute seldom appears in harmless packages. \\
\textbf{File-related features.} Research conducted earlier in the NPM ecosystem \cite{scalco2022feasibility} demonstrated that during typosquatting and combosquatting attacks, the malicious package imitated a popular package by utilizing the README file. Therefore, we resort to examine files other than .py files, such as PKG-INFO and setup.cfg files. Subsequently, we formulate the hypothesis that these files might play a role in discerning between malicious and benign packages. From this perspective, we extract two distinct features. \\
\underline{\textit{Has minimum setup configuration:}}
In the last few years, package distribution guidelines are constantly evolving \cite {vaidya2019security, reitz2016hitchhiker}. 
In Python development, both setup.cfg and setup.py are common for the purpose of packaging and distribution. 
The setup.cfg file includes the configuration of various aspects of package distribution, such as metadata details. This approach is considered cleaner, more contemporary, and modular, enabling efficient management of settings without cluttering the main script. We compute this feature by parsing the content of the setup.cfg file.  We define the minimal configuration as equivalent to the default contents automatically generated by the utilized packaging tool. We hypothesize that attackers focus on the malicious payload rather than the package design. In some instances, attackers may leave the setup.cfg file untouched, adhering to a minimal configuration. Hence, we assume that the absence of a robust design is posited as a potential indicator of a malicious package, particularly when the setup.cfg contents resemble the minimum configuration.
To support our assumption, we examine our dataset, discovering that 79\% of the malicious packages exhibit inadequate design, while only 45\% of benign packages deviate from the best practices.




% ...

\noindent
\underline{\textit{Has mismatch license:}}  A grey literature \cite {licenses} highlighted that around 10\% of PyPI packages lack clear usage licenses, posing a potential risk for malicious attacks. Python packages employ three methods to express licenses: license classifier, field, and file \cite{xu2023understanding, licenses}. We calculate this feature by examining the agreement among the types of licenses employed in the aforementioned three methods. Trusted packages adhere to proper license usage, as indicated by a recent study\cite {chinthanet2021makes}. However, our dataset analysis reveals that approximately 66\% of malicious packages exhibit license discrepancies, compared to just 1\% in benign packages.

\par
Until this point, we have prepared all the features for both malicious and benign packages. Next, we perform pairwise Pearson correlation \cite{bollen1981pearson} between the features to check whether two independent variables have a linear relationship. We found no correlation.

\subsection{Classifiers and Performance Evaluation} \label{sub:evaluation}
To perform our predictions, we leverage six machine learning classifiers from Scikit-learn python library \cite{pedregosa2011scikit}:
Random Forest, Support-Vector Machine, Decision Tree,
Multilayer Perceptron, Naive Bayes (Bernoulli version), and Stacking. These classifiers have been used in prior works \cite{sejfia2022practical,ohm2022feasibility}, as well as other software engineering works \cite{golzadeh2021ground,abdalkareem2020machine,treude2016augmenting}. To measure the performance of each classifier, we compute the precision, recall, and F1-score.

\section{Case Study Results} \label{study_result}

\begin{table*}[]
     \vspace{-4mm}
     \caption{Performance of stacking classifier on a train dataset.}
    \small
    \textbf{Best configuration (bold):} num-folds = [\textbf{10}, 15] \hspace{0.5 em}
    num-trees = [\textbf{12},100, 200] \hspace{0.5 em}
    num-words =[200, \textbf{500},1000, 2000]\\
    tokenizer modes =  [\textbf{binary}, tfidf, count, freq] \hspace{0.5 em}
    lower-states = [ False, \textbf{True}] \hspace{0.5 em} (M/B) = (Malicious/Benign)  
    \centering
    \begin{adjustbox}{center}
  \small
\begin{tabular}{lllllll}
\toprule
& &\textbf{Overall} & & &\textbf{Per-class (M/B) } &  \\

 \textbf{Input Features} & \textbf{Precision (\%)} & \textbf{Recall (\%)} & \textbf{F1-score (\%)} & \textbf{Precision (\%)} & \textbf{Recall (\%)} & \textbf{F1-score (\%)}
\\
 
\midrule
 
   Metadata-Code-File & 94 & 77 & 84 &(89/99)&(55/100)& (68/99) \\
  \hline 
 Text & 97& 83& 89 &(96/99)&(67/100)& (79/100)   \\     
Text + preprocessing & 98& 83& 89 &(97/99)&(67/100)& (79/100) \\
    \hline

  Text + Metadata-Code-File & 98& 91& 94 &(96/100)&(82/100)& (88/100)\\
 
 \textbf{Text + Metadata-Code-File  + preprocessing}   & \textbf{98}& \textbf{91}& \textbf{94} & \textbf{(97/100)}& \textbf{(81/100)}& \textbf{(89/100)} \\ 
 

   \hline
  
\end{tabular}
 
   \label{tab:perf_rf_manipulation}
   \end{adjustbox}
\end{table*}





 \subsection{RQ1: How accurate is our approach in classifying malicious packages?}

\label{sub:rq2}
\noindent
\textbf{Motivation:}
Most registries have little to no review process for publishing packages \cite{duan2020towards}, which can be exploited by attackers to publish different types of malware to harm all downstream stakeholders. To maintain the ecosystem health from malicious actions, we need to identify the malicious packages among millions of published packages. Furthermore, different scanning methods have been developed to detect packages that raise suspicion rather than those that are undoubtedly malicious. In this RQ, we aim to evaluate the effectiveness of our approach in detecting malicious packages. Our approach helps registry maintainers to have a timely and accurate prediction of the
malicious packages even before publishing them, and keeps the registry clean.

\noindent
\textbf{Approach:} To answer this question, we evaluated the classifiers in predicting malicious packages in the two datasets discussed in Section \ref{sub:feature_selection}. 
During training and validation, we performed a stratified 10-fold cross-validation, which trained six classifiers on 90\% of the dataset and measured precision, recall, and F1-score on the remaining 10\%. 
To ensure that the results are not skewed by a particular random initialization or split, this process is repeated ten times for each classifier, and then an average performance of these runs is computed to present the overall performance for that classifier.
Moreover, to understand the sort of vocabulary used in distinguishing malicious packages from benign ones, we extracted all tokens from a text feature and converted them to lowercase. We removed stop words from the set of tokens, and we evaluated the impact of these choices on the performance of classifiers. Then we used a combination of standard NLP techniques, including keywords and stop-word removal to convert the text into tokens. Then we fed these tokens as inputs for the text classification algorithms with/without the feature set extracted from the metadata-related, code-related, and file-related features.
We relied on  \textbf{Scikit-learn} python library \cite{pedregosa2011scikit} for all classifiers, with the best configurations as shown in Table \ref{tab:classifiers}, and \textbf{Keras} python library \cite{chollet2018keras} for text prepossessing (Tokenizer).
Next, to understand the generalizability of our model, we tested our trained model on the test set discussed in Section~\ref{sub:collecting-PyPI-libraries}, applying the same process shown in Figure \ref{fig:workflow} to extract the corresponding features. For this evaluation, we reported only results from our best model, the Stacking classifier. 


 

\begin{table}
    \centering
    \caption{Performance of different classifiers under precision (P), recall (R), and F1 score.}
    \begin{adjustbox}{width=\columnwidth,center}
    
\begin{tabular}{llll}

\toprule
   \textbf{Classifier} & \textbf{P (\%)} & \textbf{R (\%)} & 
   \textbf{F1 (\%)} \\
\midrule
    Random Forest (RF) [n\textunderscore estimators=12]  &97      &83     & 89 \\
    Support Vector Machine (SVM) [kernel= linear] & 99      &80     &87 \\
    Decision Tree (DT) [max\textunderscore depth=15] & 89     &89     &89 \\
    Multilayer Perceptron (MLP) [activation= logistic, solver= adam] & 97      &90     &93 \\
    Naive Bayes-Bernoulli (NB) &54 &81 &52 \\
    \hline
    \textbf{Stacking (RF+SVM+MLP+DT+NB)} & \textbf{98} & \textbf{91} & \textbf{94}\\  
\bottomrule
\end{tabular}

    \label{tab:classifiers}
    \end{adjustbox}
\end{table}

\noindent
\textbf{Fitness results:} Table \ref{tab:classifiers} presents the performance of experimented classifiers in terms of precision, recall, and F1-score values. Generally, the classifiers achieve high performance (F1-score > 87\%) in identifying malicious packages in all classifiers except the Naive Bayes (52\%). Ensemble stacking classifier outperforms all other classifiers achieving the highest F1-score (94.2\%) considering text vocabulary and metadata-related, file-related, and code-related features. Upon closer examination of the impact of using text vocabulary for the best performed classifier (i.e., stacking), the results show that the text vocabulary has a significant contribution to the prediction. When comparing the performance of stacking classifier only on metadata-related, file-related, and code-related as input features, we find that while the combination of most extracted features (i.e., metadata-related, file-related, and code-related features) achieve F1-score (84\%) , text vocabulary feature achieves higher F1-score (89\%) as shown in Table \ref{tab:perf_rf_manipulation}. The evaluation assures our conjecture that text vocabulary could be used as a strong candidate feature to distinguish malicious 
package from benign one.

    
     


    
     

\begin{table}
    \centering
 \caption{Performance of our approach on the training and test dataset.}
     \small
     \begin{adjustbox}{width=\columnwidth,center}
    \begin{tabular}{lll|ll|llll}
\toprule
\textbf{Dataset} &\textbf{Package type} & \textbf{ Total} & \textbf{FP} & \textbf{FN }& \textbf{Precision} & \textbf{Recall}& \textbf{F1-score } & \\
&&&&&(\%)& (\%)& (\%)&\\

\hline
Training &popular & 5,193 &3&-&  98 &91&94& \\
&malicious &  138 & -&23&   &  &&  \\
\hline
Test &typical & 397 & 0&-&96  & 87& 90&  \\

& malicious & 143  &-&38& &  &  &

\end{tabular}


\label{tab:performance_tarin_test}
\end{adjustbox}
\vspace{-4mm}
\end{table}

\noindent
\textbf{Test results:}
To test the generalizability of our trained model, we report the classification performance when inferring only the test set (the model was not trained further). 
The results shown in Table \ref{tab:performance_tarin_test} display both the instances of false positives and false negatives generated by our approach when employed on the test set. 
We observe that our approach performs consistently well, reaching a F1-score of 90\%, suggesting that our trained model is capable of identifying malicious packages even when regular (non-popular) Python packages are in the mix.
We can observe from Table \ref{tab:performance_tarin_test} that all typical packages are correctly classified as benign packages. 
On the other hand, when looking at the malicious packages, we observe that 105 out of 143 are classified correctly as malicious packages. 
However, the 38 malicious packages that were wrongly classified as benign were clones of popular packages, to inject their malicious payload. 
Upon manual analysis, we note some of these package's malicious payloads seem insufficient for our model to distinguish them from their benign counterparts. Further exploration of other features may help reduce these false negative mistakes. 
\begin{tcolorbox}

\textbf{The stacking ensemble classifier performs best in detecting malicious packages, with a fitting F1-score of 94.2\% and a test F1-score of 90\% on unseen data. The text vocabulary features contribute significantly to distinguishing malicious from benign packages.}
\end{tcolorbox}


  

\subsection{RQ2: What features are the best indicators of malicious packages?}
\label{sub:rq2}
\noindent
\textbf{Motivation:} Given that stacking classifier achieves a high fitting and test performance in identifying the malicious packages, we want to better
understand the most important features that contribute to the prediction. By knowing these important features, we can set the characteristics that distinguish malicious packages from benign ones.

\begin{figure} 
\includegraphics[width=\columnwidth]{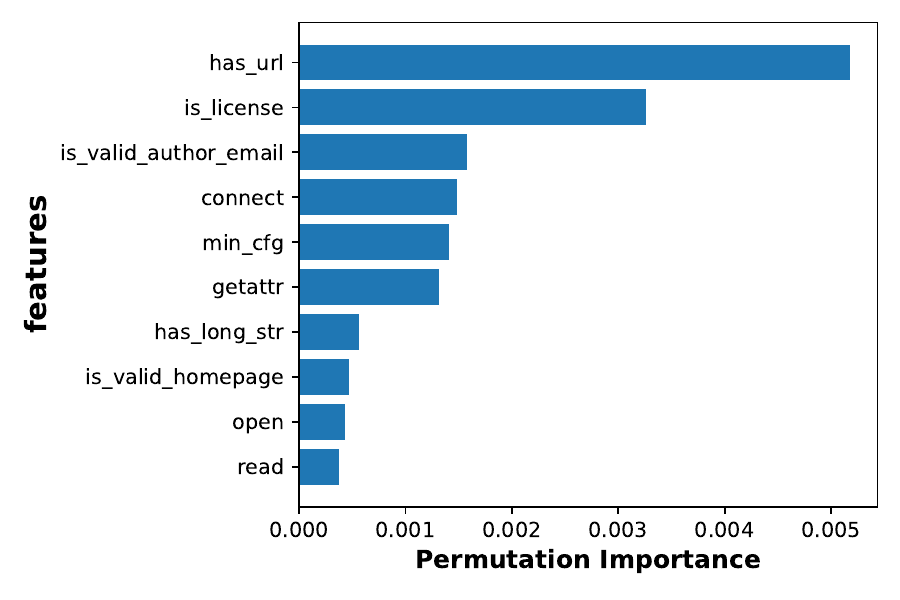}
\caption{The permutation feature importance.}
\vspace{-5mm}
\label{fig:permut_feature}
\end{figure} 

\noindent
\textbf{Approach:} We rely on the permutation feature importance technique \cite{altmann2010permutation}
to find the most useful features in the stacking classifier. 
This technique randomly permutes the values of one feature
while preserving the values of the remaining features.
This process is applied to
all features discussed in Section \ref{sub:feature_selection}. We use permutation\textunderscore importance function in the Scikit-learn library \cite{pedregosa2011scikit} to compute the feature importance values of the stacking classifier.



\noindent
\textbf{Results:} We find that the most important features are related to the following features: has suspicious url, has a licence, has a valid author email, has minimum configuration, has long string, and has a set of suspicious API (\textsc{getattr, connect, open, and read}), as shown in Figure \ref{fig:permut_feature}.
Upon a deeper examination of the percentage of the most important feature in  malicious and benign packages in our dataset, we find that while 28\% of malicious packages contain a suspicious url, only 9\% of the benign packages include it.
Another important feature is the "Has a License." Based on our analysis, the result shows that 77\% of the malicious packages have a mismatch license between the three known locations (i.e., file, classifier, and field), and only 1\% of the benign packages have this property. This confirms that these features have a significant impact on the target variable, at least in the given model's context. Moreover, it is evident that features such as "Has long string" and "Has suspicious URL" possess attributes indicative of malicious packages.
Moreover, our dataset suggests that malicious packages tend to have more invalid homepage (80\% compared to 27\% for benign packages). The same observation for the "Has minimum configuration" (79\% compared to 45\% for benign packages).
Note that only the suspicious APIs are not enough to capture most malicious packages (low recall) as shown in Table \ref{tab:classifiers}, but when combined with metadata features, the model reaches excellent performances.

\begin{tcolorbox}
\textbf{Code-related, file-related, and metadata-related features are all contribute to identify the malicious packages. Moreover, suspicious APIs such as getattr, connect, open, and read, contribute to distinguishing malicious from benign packages.  }  
\end{tcolorbox}

\subsection{ RQ3: Is our approach useful?}
\textbf{Motivation:} In RQ1, we assessed the performance of our approach and found that the stacking classifier achieves the best results in identifying malicious packages with F1 score of 94.2\%. Moreover, a recent study \cite{vu2023bad} has brought attention to the fact that popular packages are different from
a typical Python package. The study emphasizes that these popular packages demonstrate enhanced engineering and a stronger alignment with standard Python programming conventions. Consequently, using only popular packages as the benign
dataset might lead to unrealistic benchmark results since these packages might be relatively easy for detection tools to classify as benign.

\noindent
\textbf{Approach:} To put our results into perspective, we conducted two experiments. We compared our approach with (1) two benchmarking tools: Bandit and Packj, as in prior works  \cite{vu2020towards, vu2022benchmark, sejfia2022practical}, (2) the method of \citet{ohm2022feasibility}.
For the first experiment, we utilized the above mentioned tools for the reasons specified in Section \ref {malicious}. We selected these tools, despite their differing threat models, to explore their effectiveness in detecting malicious dependencies and to measure the manual effort needed to filter through all signals they generate. We run both tools and manually examine the returned alerts to identify malicious packages, simulating how practitioners use these tools to identify suspicious functions and conclude the presence of malicious packages in their applications.
We avoid the computational overhead associated with Packj tool by focusing solely on the static code analysis capability for identifying API usage, which is then processed further to isolate the relevant phantom lines of code that constitute the text-related feature.
\noindent
In this experiment, we determine the number of alerts generated by applying the bandit and packj tools on a random subset containing 50 benign packages and another 50 samples of malicious packages sourced from the test dataset (Table \ref{tab:performance_tarin_test}). 
This evaluation is conducted in two situations: the whole package and exclusively the setup.py file.\\ 
In the second experiment, we exploited the method proposed by \citet{ohm2022feasibility}, due to its close similarity to ours. Their dataset, drawn from the same source as ours, comprised nearly equivalent sizes, with 150 malicious npm packages compared to our 138 Python packages. 
We leveraged their method, which relied on the intersection of outcomes from three classifiers (SVM, RF, and MLP).
It is important to highlight, however, that the approach of Ohm et al~\cite{ohm2022feasibility} was tailored to the JavaScript ecosystem, while our approach accounted for the specificities of Python packages.
Thus, in this experiment, we are also evaluating how ecosystem-tailored features contribute to classifying malicious Python packages.

\begin{figure*}[]
   
    \centering
   \includegraphics[height=0.2\textheight,width=0.85\textwidth ]{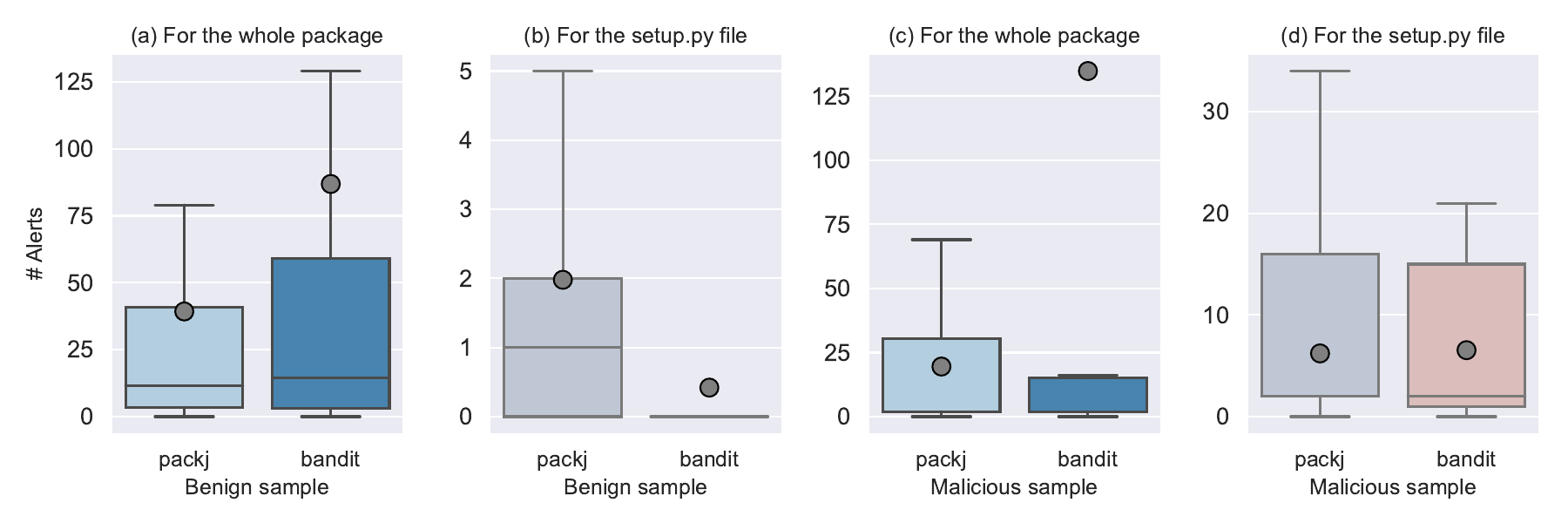}
   
     \caption{The average number of alerts generated by bandit and packj tools from the whole package and setup.py file.}
      \vspace{-2mm}
      \label{alerts}
\end{figure*}

\begin{table}
    \caption{Number of generated alerts for a random sample of external dataset for packj and bandit tools.}
    \centering
    \begin{adjustbox}{width=\columnwidth,center}
    \small
   
    \begin{tabular}{ll|ll|ll}
\toprule
\textbf{Malicious Package} &  & \textbf{\# Packj Alerts}&  & \textbf{\# Bandit Alerts} & \\

         &      &     whole pkg & setup& whole pkg & setup  \\

\bottomrule

2022-requests-3.0.0   && 69&	16	&555	&21\\
typing-unions-3.10.0.1&&	38&	0&	61&	1\\
rumihelling-0.0.1&&	15	&0&	4&	0\\

dlcsord-1.0.3&&	0&	0&	0&	0 \\
\hline
\end{tabular}

     \label{tab:comparison2}
     \end{adjustbox}
   \vspace{-6mm}
\end{table}

\textbf{Result:} Figure \ref{alerts} presents the average number of alerts generated by bandit and packj tools in two scenarios, whole package and only setup.py file. The presented findings include both true positives and false positives.
The result of Figure \ref{alerts}  (a) reveals that both the bandit and packj tools generated a considerable number of false alerts when assessing the entire benign package. Nonetheless, by considering only the setup.py file, Figure \ref{alerts}  (b), the average alert count peaked at a maximum of two alerts.
Clearly, the number of alerts, for both tools has risen when considering the whole package both of benign and malicious packages, Figure \ref{alerts}  (a) and (c). 
\noindent
Moreover, it can be noted  from the same figure that when focusing solely on the setup.py file, we see that malicious packages tend to produce a higher number of alerts, averaging around six alerts, Figure \ref{alerts}  (d) , in contrast to benign packages (two alerts as in Figure \ref{alerts}  (b)). This reconfirms the earlier conclusions presented in references such as \cite{ohm2020backstabber, vu2020towards}, which identified that the setup.py file was the most targeted file by attackers.



     

\begin{figure}
    \centering
   \includegraphics[height=0.2\textheight,width=\columnwidth]{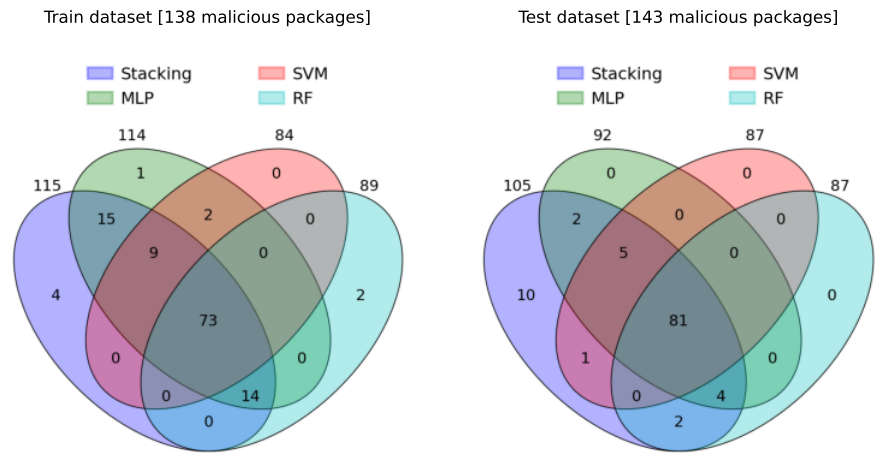}
        \caption{The result of the intersection approach \cite{ohm2022feasibility} of three classifiers on our train and test dataset.}  
      \label{venn}
      \vspace{-6 mm}
\end{figure}

To provide further details, Table \ref{tab:comparison2} shows the number of alerts produced by the packj and bandit tools for a subset of a selected sample of malicious packages . We observed a significant volume of alerts being generated by both tools, especially when examining the complete package scenario. This necessitated significant manual effort from developers to verify the status of these packages. The bandit tool, in particular, generated 555 alerts, and a manual investigation revealed that a number of these alerts were wrongly triggered. Our approach misclassified 16 samples out of 50 (32\%)  where the total misclassificaton was 38 out of 143 samples (about 27\%) as shown in Table \ref{tab:performance_tarin_test}. This highlights the capability of our approach to reduce the manual workload for registry maintainers.
In evaluating the two tools for package security, it is evident that both tools have limitations in identifying suspicious packages, as shown in instances like typing-unions-3.10.0.1, rumihelling-0.0.1, and dlcsord-1.0.3 (Table \ref{tab:comparison2}), particularly when focusing solely on the setup.py file. Notably, our approach successfully classified all benign packages, in contrast to packj and bandit tools, which generated excessive false alerts for benign packages. Nonetheless, it's worth emphasizing that our method results in a relatively small number of false negatives. 
\noindent
In the second experiment, we compared our approach, with features tailored to the Python ecosystem, against the approach of \citet{ohm2022feasibility}. 
We observed that the method proposed by \citet{ohm2022feasibility} failed to identify certain malicious packages. In contrast, our approach proved to be more effective, successfully detecting a wider range of these packages, as depicted in Figure \ref{venn}. When employing a stacking classifier, we managed to detect 115 out of 138 in the training dataset and 105 out of 143 in the testing dataset. However, relying on the intersection method of the three classifiers only enabled us to detect 73 out of 138 and 81 out of 143 for the training and testing datasets, respectively. 
This indicates that our approach, which is tailored to the Python ecosystem, is more effective in detecting malicious Python packages compared to the approach proposed by Ohm et al.~\cite{ohm2022feasibility}.

\begin{tcolorbox}
\textbf{Our approach is reasonably precise and does not produce an overwhelming number of false negatives and false positives, making manual investigation extremely feasible, when apply to the entire ecosystem.}  
\end{tcolorbox}

\section{Model Misclassifications}\label{discussion}

This section analyzes the misclassification of packages by our model in both training and external datasets to understand the underlying reasons. In the training dataset, 23 out of 138 malicious and 3 out of 5,193 benign packages were misclassified. For the external dataset, 38 out of 143 malicious packages were misclassified, with no benign misclassifications among 397.

\noindent
\textbf{Misclassification in popular packages.}
Our investigation found that less than 0.06\% of popular packages were misclassified by our model, primarily due to poorly structured code and deviation from standard Python conventions. Misclassification occurred because of a lack of essential metadata, such as homepage, author email, or license, as seen in packages like ordereddict-1.1, cookies-2.2.1, and blob-0.16. For instance, cookies-2.2.1 had an unknown license and used suspicious APIs like 'getattr', 'setattr', 'open', and 'call', which can indicate malicious behavior. We merged vocabulary from the setup.py file with lines from suspicious files, highlighting how missing metadata and the presence of suspicious APIs lead to misclassification. Ultimately, misclassifying benign packages as malicious is considered less risky than the reverse scenario.

\noindent
\textbf{Misclassification in malicious packages.}
Our model falsely classified some packages due to many reasons.
(1) The package incorporates a reference to an external harmful function. For examaple, libpesh-0.1 package has been identified as malicious because it includes a reference to a harmful function called "rn" which can be found in the file named \textsc{entry\textunderscore point.txt} as "\textsc{eggsecutable = libari.pr:rn}." Due to the absence of the text code, which are critical feature that our model relies on, this incident has the potential to deceive the model. 
(2) Using a group of suspicious APIs that are commonly employed by benign packages. For example, the package bzip-0.98 has been designed to appear as the legitimate bz2file package, and its installation script, \textsc{setup.py} has been altered to contain a malicious code that is not particularly harmful.

\section{Threats to validity}\label{threats}
\noindent
\textbf{Internal Validity.} The internal validity threats in this study include the potential for false positives and false negatives, although these are manageable. Manual review helps mitigate false positives. The dataset may also be biased due to clustering of similar malicious samples and assuming popular PyPI packages as benign. The selection criteria for popular packages could introduce variability, impacting the results. Despite these threats, we believe the dataset is sufficient for the experiments.

\noindent
\textbf{Construct Validity.}
One threat to construct validity is the reliance on the Packj tool \cite{packj-vetting-tool} for static analysis, as we used it primarily to extract features for training our text classifier rather than drawing final conclusions. Another threat arises from the feature set construction, as we inherited some features from prior studies that effectively identify malicious packages \cite{sejfia2022practical, garrett2019detecting, packj-vetting-tool, duan2020towards}. However, this reliance may limit our outcomes, highlighting the need for future research to investigate additional features and their impact.

\noindent
\textbf{External Validity.} 
External validity concerns the generalization of our findings. In our study, we assessed the performance of our approach using 5,193 bening packages and 138 malicious ones. Hence, our results may not generalize to other datasets, as malicious dataset may not accurately represent all malicious packages in the wild, and may there are  malware with different characteristics than those in our dataset. However, we still believe that our dataset is comprehensive and serve the goal of this study. Moreover, to alleviate this threat, we evaluate the model on an external (unseen) dataset, and we found that our approach
performs very well, suggesting its generalizability.

\section{Conclusion}\label{conclusion}
We presented a machine learning-based method for detecting malware in PyPI packages using features from text, file, code, and metadata. Among the six classifiers evaluated, the stacking classifier performed best, while Naive Bayesian failed to detect known malicious packages. Our approach demonstrated practical effectiveness with low false negatives on external datasets. This method holds promise for automatic malware detection in PyPI. Future work includes expanding features and applying the technique to other ecosystems like NPM and RubyGems.

 \bibliographystyle{ACM-Reference-Format}

\bibliography{bibliography}


\end{document}